\documentclass[prl,10pt,twocolumn,nofootinbib,preprintnumbers,amssymb,amsfonts,amsmath,superscriptaddress,showpacs,hyperref]{revtex4-1}

\usepackage{bm}
\usepackage{url}
\usepackage{graphicx}
\usepackage{multirow}
\usepackage{epstopdf}
\usepackage{subfigure}

\def\apj{ApJ}
\def\apjl{ApJL}

\def\apss{Ap\&SS}               
\def\planss{P\&SS}              

\def\jgr{Journal of Geophysical Research}

\def\gtsima{$\; \buildrel > \over \sim \;$}
\def\ltsima{$\; \buildrel < \over \sim \;$}
\def\gsim{\lower.5ex\hbox{\gtsima}}
\def\lsim{\lower.5ex\hbox{\ltsima}}
\def\simleq{\; \raise0.3ex\hbox{$<$\kern-0.75em \raise-1.1ex\hbox{$\sim$}}\; }

\newcommand{\GeV}{{\rm GeV}}

\newcommand{\kpc}{{\rm kpc}}

\newcommand{\cm}{{\rm cm}}
\newcommand{\km}{{\rm km}}

\newcommand{\s}{{\rm s}}

\newcommand{\AU}{{\rm AU}}
\newcommand{\nT}{{\rm nT}}
\newcommand{\GV}{{\rm GV}}

\newcommand{\dragon}{{\sc Dragon}}

\newcommand{\SolarProp}{{\sc HelioProp}}

\begin{document}

\title{Low energy cosmic ray positron fraction explained by charge-sign dependent solar modulation}

\author{Luca Maccione}
\affiliation{Ludwig-Maximilians-Universit\"at, Arnold Sommerfeld Center, Theresienstra{\ss}e 37, D-80333 M\"unchen}
\affiliation{Max-Planck-Institut f\"ur Physik (Werner-Heisenberg-Institut),
F\"ohringer Ring 6, D-80805 M\"unchen}


\date{\today}

\begin{abstract}
We compute cosmic ray (CR) nuclei, proton, antiproton, electron and
positron spectra below 1 TeV at Earth by means of a detailed
transport description in the galaxy and in the solar system. CR
spectra below 10 GeV are strongly modified by charge-sign dependent
propagation effects. These depend on the polarity of the solar
magnetic field and therefore vary with the solar cycle. The puzzling
discrepancy between the low-energy positron fraction measured by
PAMELA and AMS-01 is then easily explained by their different
data-taking epochs. We reproduce the observed spectra of CR light
nuclei within the same galactic and solar-system propagation model.
\end{abstract}

\preprint{LMU-ASC 83/12; MPP-2012-155}
\maketitle

{\em Introduction}: The propagation of Cosmic Rays (CRs) in the Galaxy is far from being fully understood. 
%
One major challenge is represented by the spectrum of the positron fraction (PF), $e^{+}/ (e^{-}+e^{+})$, measured now by PAMELA and Fermi with high accuracy at energies ranging from below 1 GeV up to about 100 GeV \cite{Adriani:2008zr,FermiLAT:2011ab,Abdo:2009zk,Ackermann:2010ij}, and previously by AMS-01 \cite{Aguilar:2007yf} and many other experiments. While the well known increase of the PF with increasing energy above 10 GeV has attracted most of the attention of the astroparticle physics community, the disagreement between PAMELA and AMS-01 PF below 10 GeV has motivated several investigations \cite{Delahaye:2008ua,GastSchael,DellaTorre:2012zz,Delahaye:2010ji} but has never been convincingly explained so far. 

Before they are detected at Earth, CR $e^{-}$ and $e^{+}$ lose energy due to solar winds while diffusing in the solar system \cite{Gleeson_1968ApJ}. This modulation effect depends, via drifts in the large scale gradients of the solar magnetic field (SMF), on the particle's charge including its sign \cite{1996ApJ...464..507C}. Therefore, it depends on the polarity of the SMF, which changes periodically every $\sim$11 years \cite{wilcox}. Besides the 11 year reversals, the SMF has also opposite polarities in the northern and southern hemispheres: at the interface between opposite polarity regions, where the intensity of the SMF is null, a heliospheric current sheet (HCS) is formed (see e.g.~\cite{1981JGR....86.8893B}). The HCS swings then in a region whose angular extension is described phenomenologically by the tilt angle $\alpha$. The magnitude of $\alpha$ depends on solar activity. Since particles crossing the HCS suffer from additional drifts because of the different orientation of the magnetic field lines, the intensity of the modulation depends on the extension of the HCS. The PF spectra measured by AMS-01 and PAMELA can therefore differ because they were measured in different periods of the solar activity \cite{GastSchael,DellaTorre:2012zz,Strauss:2012zza}. 

\begin{figure}[tbp]
\begin{center}
\includegraphics[width=0.5\textwidth]{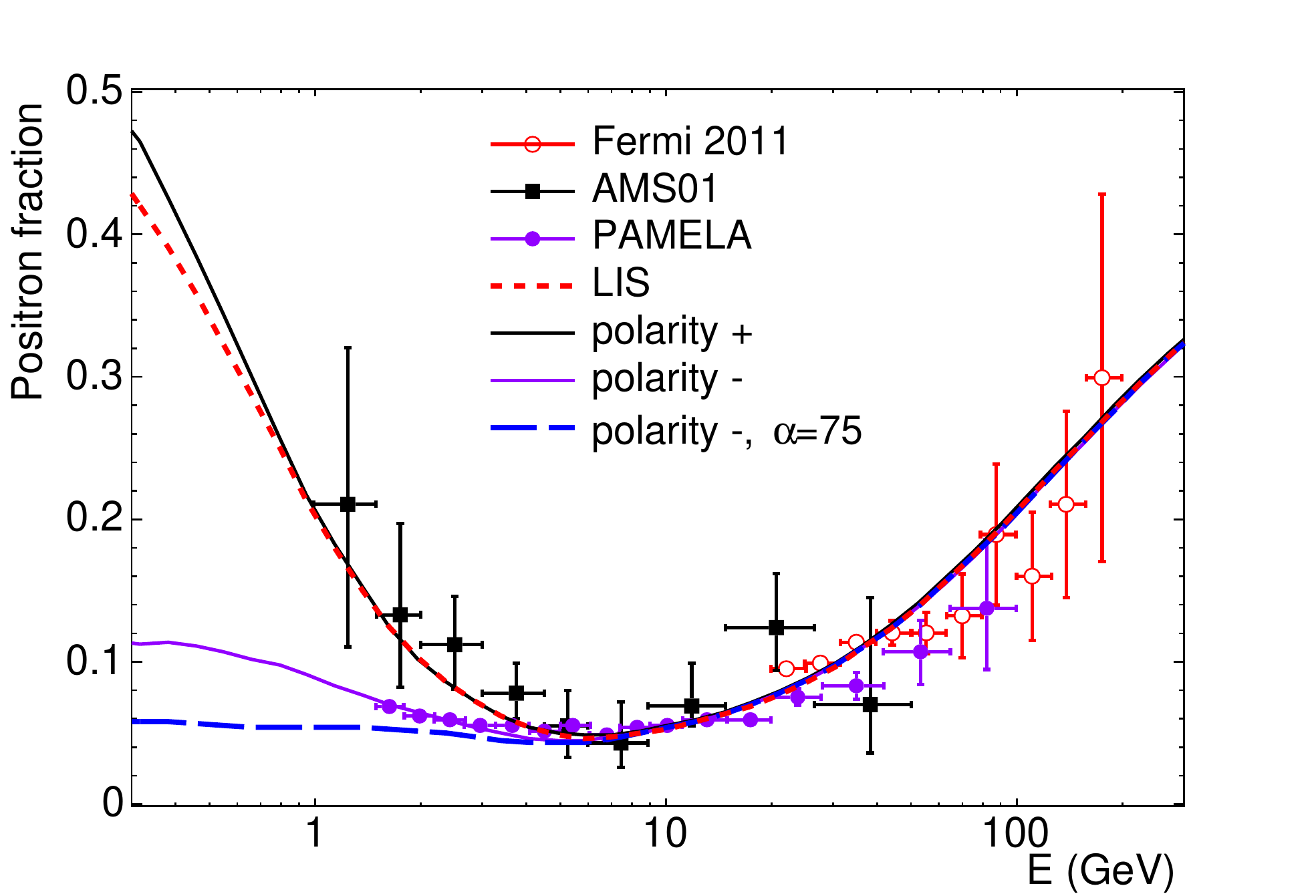}
\caption{The positron fraction measured by Fermi, PAMELA and AMS-01 is shown. The LIS is shown as the red dashed curve. Solid curves show the Earth positron fraction computed evolving the LIS for $\alpha=30^{\circ}$ and positive polarity (black) or negative polarity (violet). The long-dashed blue curve represents a prediction for AMS-02, which is taking data in a period of negative polarity and high solar activity.}
\label{fig:PF}
\end{center}
\end{figure}

We show in Fig.~\ref{fig:PF} how changing polarity affects the PF, computed using the methods that will be described later.
These effects are clearly evident below 10 GeV in the model curves and are large enough to explain the discrepancy between AMS-01 and PAMELA data, which were taken in periods of opposite polarity and comparable tilt angle $\alpha\simeq 30^{\circ}$. Indeed, we reproduce very well the PF measured by AMS-01 with the solar model adapted to AMS-01 data taking conditions, but remarkably we also nicely reproduce PAMELA data by tuning the solar model to the appropriate PAMELA period. This confirms that the low energy PF can be interpreted self-consistently by taking into account drift effects in solar propagation. We also show the PF computed for $\alpha=75^{\circ}$, appropriate for year 2012, during which AMS-02 has been taking data \cite{ams02}. This prediction will soon be tested by that experiment. 

This result is the main achievement of this Letter. We will support it by describing the computational methods and assumptions we adopted to compute the PF. The main novelty with respect to previous investigations \cite{GastSchael,DellaTorre:2012zz} is that for the first time we have computed the PF within a galactic+solar propagation model self-consistently derived from observations of CR nuclear and lepton spectra. Our results demonstrate that having a precise account of solar effects, going beyond the force-field model of \cite{Gleeson_1968ApJ}, is crucial to interpret the data available nowadays.

{\em Method}: 
We first select one propagation model by imposing that CR nuclear spectra at Earth be correctly reproduced. We compute the galactic interstellar spectra (LIS) first and then apply a detailed 4D (3 spatial + energy) solar propagation model to CR nuclei to obtain the Earth spectrum to be compared with observations. Then, within the propagation model determined in this way, we compute the $e^{+}$ and $e^{-}$ spectra and the PF. 

Both propagation regimes are described by a general diffusion-convection-reacceleration-energy-loss equation \cite{1964ocr..book.....G,1965P&SS...13....9P}. 
We propagate CR species in the Galaxy starting from $Z=28$ with the \dragon\ code \cite{dragonweb}. To avoid boundary effects, we set the vertical boundary of the numerical domain $L=2z_{t}$, with $z_{t}$ the half-height of the diffusion region. We assume a diffusion coefficient in the Galaxy $D(r,z,p)=D_{0}\beta^{\eta}(\rho/\rho_{0})^{\delta}\exp(z/z_{t})$, where $\rho(p)$ is particle rigidity, $\beta$ is particle's velocity, $\rho_{0}=3~\GV$, $\eta$ and $\delta$ are constant parameters. For reacceleration, we use the standard description in terms of momentum diffusion, with $D_{pp} \propto p^{2}v_{A}^{2}/D$, where $v_{A}$ is the Alfv\'en velocity. Convective effects due to stellar winds are described by a convection velocity directed  along $\hat{z}$ outside the galactic plane $v_{C}(z) =v_{C0}+v_{C1}(|z|/1~\kpc)$. 

We set $\delta=0.5$, $z_{t}=4~\kpc$, $D_{0}=2.5\times10^{28}~\cm^{2}/\s$, $v_{A}=18~\km/\s$, $v_{C0}=0$ and $v_{C1}=5~\km/\s/\kpc$, $\eta=1$. We assume for all nuclei a power-law injection spectrum with spectral index $\gamma=2.3$. For primary electrons, we assume a broken power-law with $\gamma=1.5/2.65$ below/above 6 GeV. The presence of such a break in the primary $e^{-}$ spectrum is also confirmed by the analysis of the diffuse galactic synchrotron emission \cite{Bringmann:2011py,Strong:2011wd,DiBernardo:2012zu}. In order to reproduce high energy $e^{\pm}$ data, we also consider an extra-component of primary electrons and positrons with $\gamma_{\rm extra}=1.5$ and exponential cutoff at 1.2 TeV \cite{Grasso:2009ma,DiBernardo:2010is}. We fix its normalization so that we match the Fermi $e^{+}+e^{-}$ spectrum at 300 GeV, while the normalization of the primary electron spectrum is fixed at 33 GeV.


For the 4D propagation of CRs in the solar system we develop instead a new numeric program, \SolarProp. We follow the stochastic differential equation approach of \cite{2011ApJ...735...83S,2012Ap&SS.339..223S,2007JGRA..11208101A} (see \cite{gardiner2009stochastic,2012CoPhC.183..530K} for a general description), where the CR phase-space density is computed by sampling and averaging upon pseudo-particle trajectories. Each trajectory is in fact the result of a deterministic component related to the drifts, and of a random walk component, whose amplitude is sampled according to the local diffusion tensor. Pseudo-particles injected at the Earth position are followed backward in time during their propagation in the solar system until they reach the heliopause, where their properties are recorded (we refer the reader to \cite{2011ApJ...735...83S,2012Ap&SS.339..223S,2007JGRA..11208101A} for more details on the actual numerical scheme to be used). We propagate $10^{4}$ pseudo-particles per energy bin, which allows us to efficiently sample the Green function and to keep statistical fluctuations below 1\% \cite{2011ApJ...735...83S,2012Ap&SS.339..223S}. The LIS flux, which is effectively a boundary condition for this problem, is then used as an appropriate weight to determine the Earth spectrum. 

We specify our model for solar propagation by fixing the solar system geometry, the properties of diffusion and those of winds and drifts. For simplicity we assume spherical geometry, although some latitudinal dependence of the position of the heliopause, which we set at 100 AU, has been experimentally found (see \cite{Bobik:2011ig} and Refs. therein). 

We describe the solar system diffusion tensor by $\mathbf{K}(\rho) = {\rm diag}(K_{\|}, K_{\perp r}, K_{\perp \theta})(\rho)$, where $\|$ and $\perp$ are set with respect to the direction of the local magnetic field. We assume no diffusion in the $\perp\varphi$ direction and we describe as drifts the effect of possible antisymmetric components in $\mathbf{K}$. For the parallel CR mean-free-path we take $\lambda_{\|} = \lambda_{0}(\rho/1~\GeV)(B/B_{\bigoplus})^{-1}$, with $\lambda_{0}=0.15~\AU$ and $B_{\bigoplus}=5~\nT$ the value of the magnetic field at Earth position, according to \cite{2011ApJ...735...83S,2012Ap&SS.339..223S}. For $\rho < 0.1~\GeV$, $\lambda_{\|}$ does not depend on rigidity. The value of $\lambda_{0}$ and the rigidity dependence of $\lambda_{\|}$ are compatible both with the measured $e^{-}$ mean-free-path (see, e.g., \cite{Droge2005532}) and with the proton mean-free-path inferred from neutron monitor counts and the solar spot number \cite{Bobik:2011ig}. We then compute $K_{\|} = \lambda_{\|}v/3$. Perpendicular diffusion is assumed to be isotropic. According to numerical simulations, we assume $\lambda_{\perp r,\theta} = 0.02\lambda_{\|}$ \cite{1999ApJ...520..204G}. 

\begin{figure}[tbp]
\begin{center}
\includegraphics[width=0.5\textwidth]{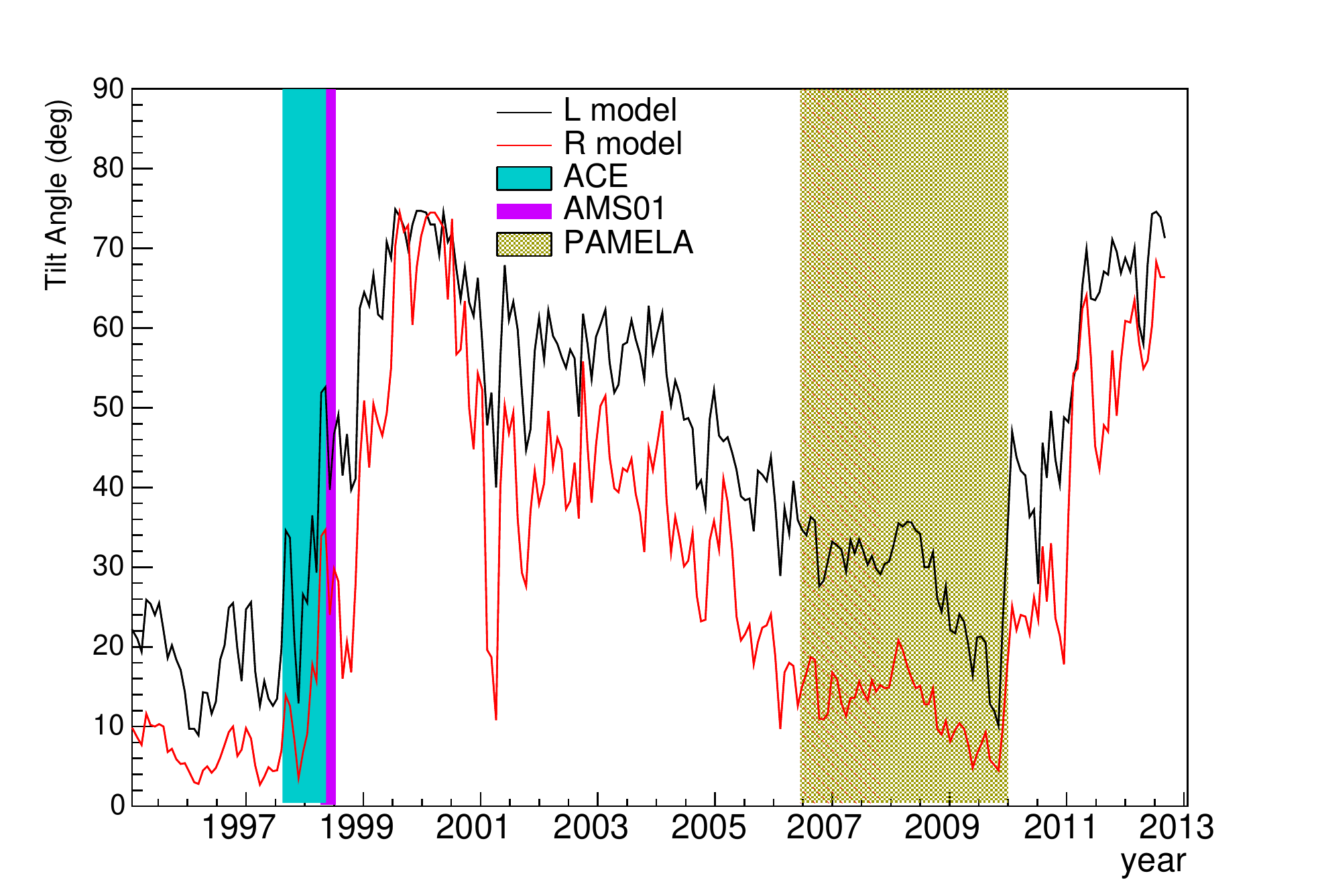}
\caption{Time evolution of the tilt angle. Vertical bands highlight data taking periods of relevant experiments. Data relevant for the positron fraction were taken by PAMELA in the period highlighted by the red area. The larger greyish shadowed area marks the data-taking period relevant for PAMELA absolute $e^{-}$ spectrum data.}
\label{fig:tiltangle}
\end{center}
\end{figure}

For the SMF, we assume a Parker spiral, although more complex geometries might be more appropriate for periods of intense activity
\begin{equation}
\vec{B} = AB_{0}\left(\frac{r}{r_{0}}\right)^{-2}\left(\hat{r} - \frac{\Omega r\sin\theta}{V_{\rm SW}}\hat{\varphi}\right)\;,
\end{equation}
where $\Omega$ is the solar differential rotation rate, $\theta$ is the colatitude, $B_{0}$ is a normalization constant such that $|B|(1~\AU)=5~\nT$ and $A=\pm H(\theta-\theta')$ determines the MF polarity through the $\pm$ sign. The presence of a HCS is taken into account in the Heaviside function $H(\theta-\theta')$. The HCS angular extent is described by the function $\theta' = \pi/2 + \sin^{-1}\left(\sin\alpha\sin(\varphi+\Omega r/V_{\rm SW})\right)$, where $0<\alpha<90^{\circ}$ is the tilt angle. The drift processes occurring due to magnetic irregularities and to the HCS are related to the antisymmetric part $K_{A}$ of the diffusion tensor as \cite{1977ApJ...213L..85J}
%
$\vec{v}_{\rm drift} = \nabla\times(K_{A}\vec{B}/|B|) = {\rm sign}(q)v/3\vec{\nabla}\times\left(r_{L}\hat{B}\right)$,
%
where $K_{A} = pv/3qB$, $r_{L}$ is the particle's Larmor radius and $q$ is its charge. We refer to \cite{2011ApJ...735...83S,2012Ap&SS.339..223S} for more details on the implementation of the HCS and of drifts. Adiabatic energy losses due to the solar wind expanding radially at $V_{\rm SW}\sim400~\km/\s$ are taken into account.

We report in Fig.~\ref{fig:tiltangle} the values of $\alpha$ \cite{wilcox} inferred from solar models, together with the periods in which relevant CR experiments were taking data.  Predictions from the ``L'' (line-of-sight boundary conditions) model seem more accurate for periods of decreasing tilt angle, while for periods of increasing $\alpha$ model ``R'' (radial boundary conditions) is more precise \cite{Ferreira2003657,2004ApJ...603..744F}.

$A$ and $\alpha$ are of particular importance for CR propagation in the heliosphere. If $q\cdot A<0$, drifts force CRs to diffuse in the region close to the HCS, which enhances their effective propagation time and therefore energy losses, while if $q\cdot A>0$ drifts pull CRs outside the HCS, where they can diffuse faster \cite{2011ApJ...735...83S,2012Ap&SS.339..223S}. 



For each CR experiment we fix the relevant solar propagation model as follows: $\alpha=30^{\circ}$ for both PAMELA and AMS-01 data, but $A=+$ for AMS-01 and $A=-$ for PAMELA. For ACE \cite{2009ApJ...698.1666G} we use $\alpha=10^{\circ}$ and $A=+$. We do not vary the intensity of the magnetic field at Earth position. 
%
%


%
\begin{figure}[tbp]
\begin{center}
\includegraphics[width=0.5\textwidth]{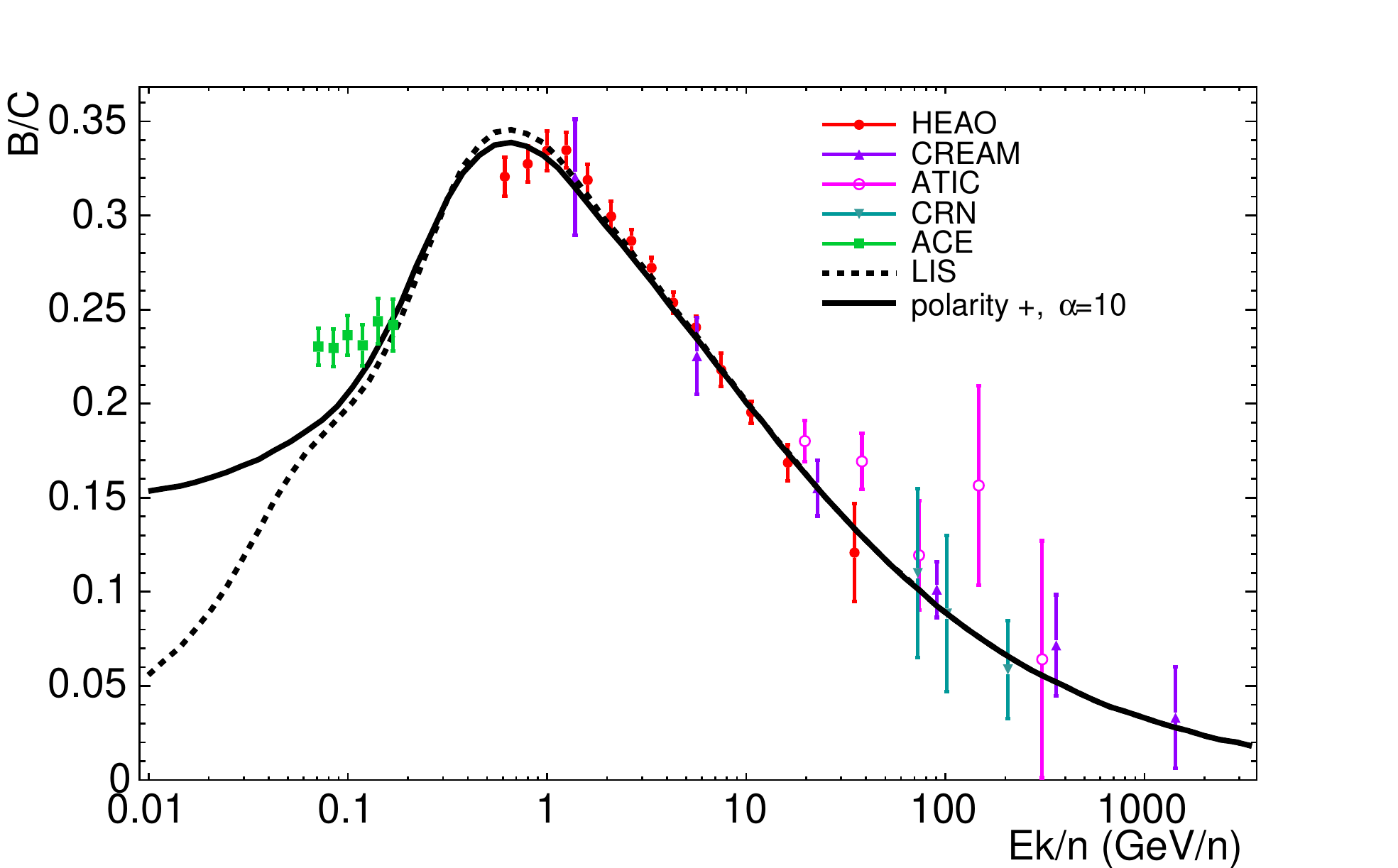}
\caption{Experimental data for the B/C ratio \cite{1989ApJ...346..997B,1990ApJ...349..625S,Panov:2007fe,Ahn:2008my} compared to our model.}
\label{fig:BC}
\end{center}
\end{figure}

\begin{figure}[tbp]
\begin{center}
\includegraphics[width=0.5\textwidth]{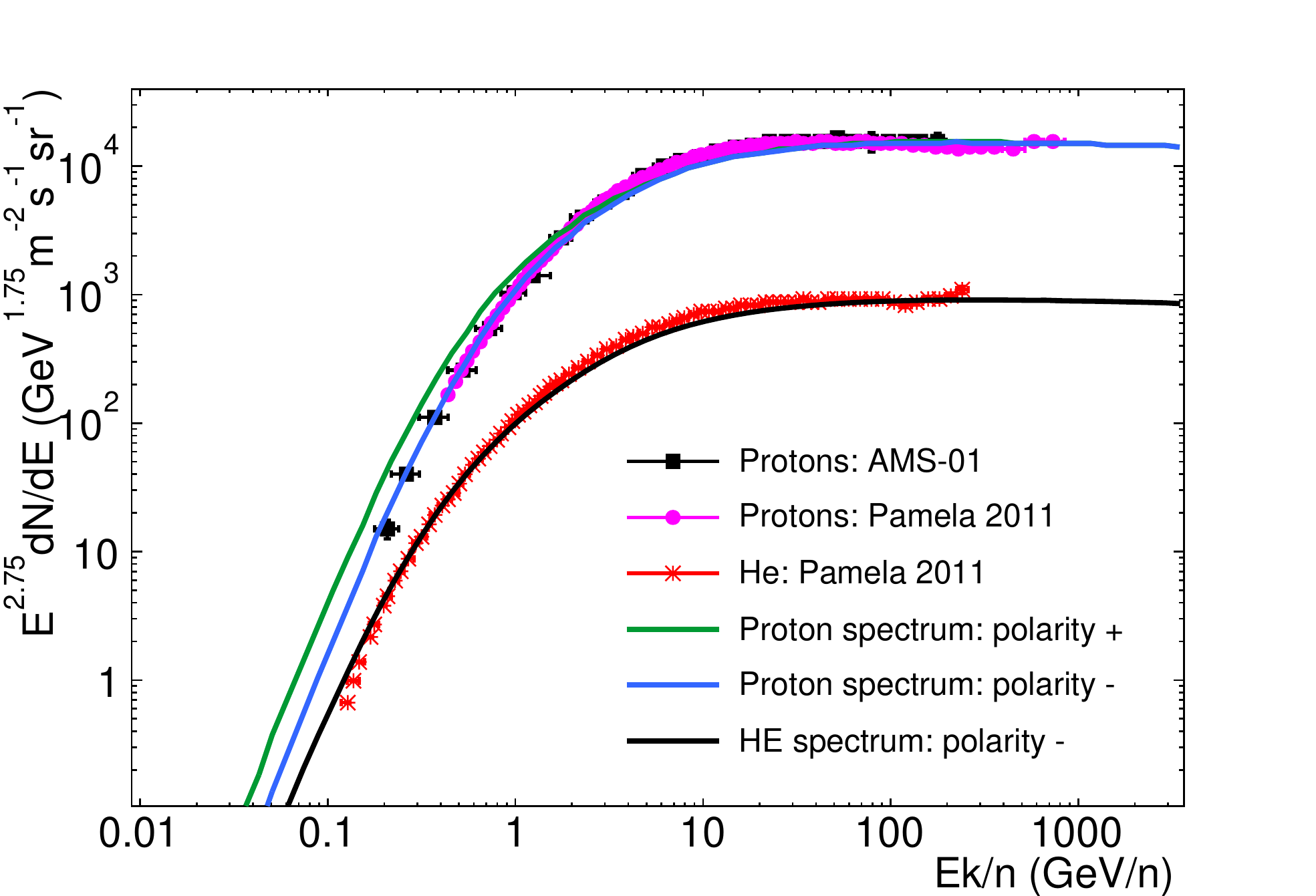}
\caption{Experimental data for proton spectra from PAMELA and AMS-01 and for the He spectrum from PAMELA compared to theoretical calculations for $\alpha=30^{\circ}$.}
\label{fig:HHe}
\end{center}
\end{figure}

\begin{figure}[tbp]
\begin{center}
\includegraphics[width=0.5\textwidth]{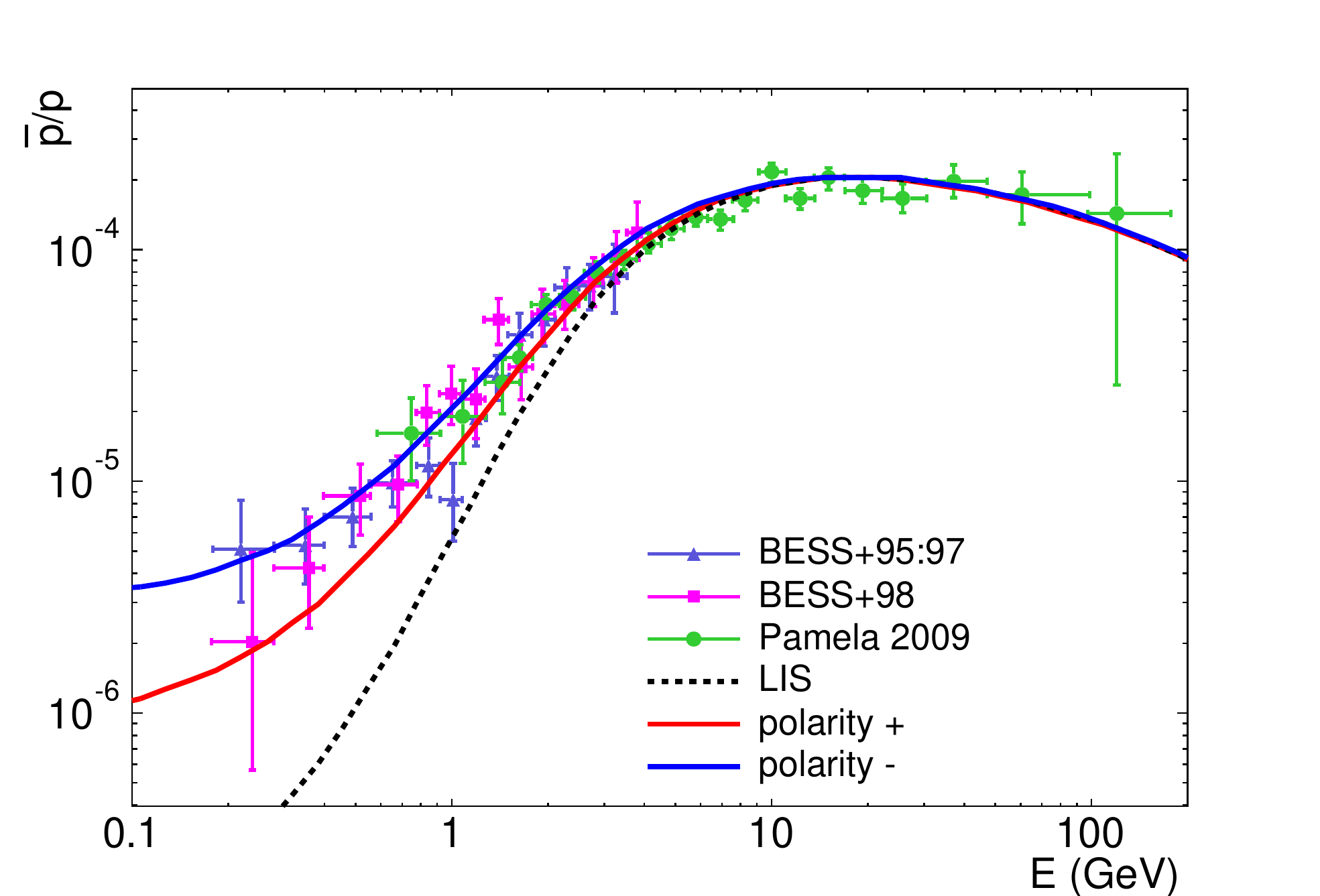}
\caption{Experimental data for the $\bar{p}/p$ ratio from PAMELA and BESS \cite{bess} compared to theoretical calculations. In order to highlight charge-sign effects, models for opposite magnetic polarity and same $\alpha=30^{\circ}$ are shown.}
\label{fig:AP}
\end{center}
\end{figure}

{\em Results}: We obtain a good description of the B/C spectrum over the entire energy range, also including ACE data \cite{2009ApJ...698.1666G}, as we show in Fig.~\ref{fig:BC}. We achieve this without resorting to large $v_{A}$, which is in conflict with the diffuse synchrotron galactic emission \cite{Strong:2011wd,DiBernardo:2012zu}, nor to modified low-energy diffusion. Combinations of these possibilities have been invoked to reproduce ACE data \cite{DiBernardo:2009ku,Putze:2010zn,Maurin:2010zp,Trotta:2010mx}. 
We checked that also the spectra of C, O and $^{10}$Be/$^{9}$Be are correctly reproduced.

%
%

The proton and He spectra measured by PAMELA are reproduced very well, without introducing a break in the injection spectrum at $\sim10~\GeV$ and with the same injection index, as shown in Fig.~\ref{fig:HHe}. Interestingly, the AMS-01 and PAMELA proton spectra are very similar at low energy, although they have been taken in different periods of solar activity. Given the observed values of $\alpha$ for the respective data-taking periods, we cannot reproduce both measurements with our model. To achieve a good match of both, we would need to assume $\alpha\simleq5^{\circ}$ for AMS-01, which is not inconsistent with observations given that the AMS-01 data taking period followed a phase of very low solar activity and that the HCS takes about 14 months to propagate from the Sun to the Heliopause \cite{Bobik:2011ig}. This has however negligible impact on the computation of secondary $e^{+}$ and $e^{-}$ in the energy range relevant for the PF, because the discrepancies are at $E\simleq 3~\GeV$. BESS spectra \cite{bess}, which have been taken over a period spanning almost a complete solar half cycle, might help understand this issue, but given that we are interested here in the comparison of the PF of PAMELA and AMS-01, we prefer to leave the study of BESS spectra for future work and compare only PAMELA and AMS-01 data. Another possibility is to more finely tune the propagation model. Indeed, by slightly adapting the propagation parameters a much better fit of both PAMELA and AMS-01 proton data can be achieved, at the price of a slight worsening of the B/C. However, we explicitly verified that our fit of the PF is not significantly affected.

Also $\bar{p}$ (not shown) and $\bar{p}/p$ spectra are very well reproduced (see Fig.~\ref{fig:AP}). In this case the separation between model calculations in different periods of solar activity is comparable to experimental errors. 

\begin{figure}[tbp]
\begin{center}
\includegraphics[width=0.5\textwidth]{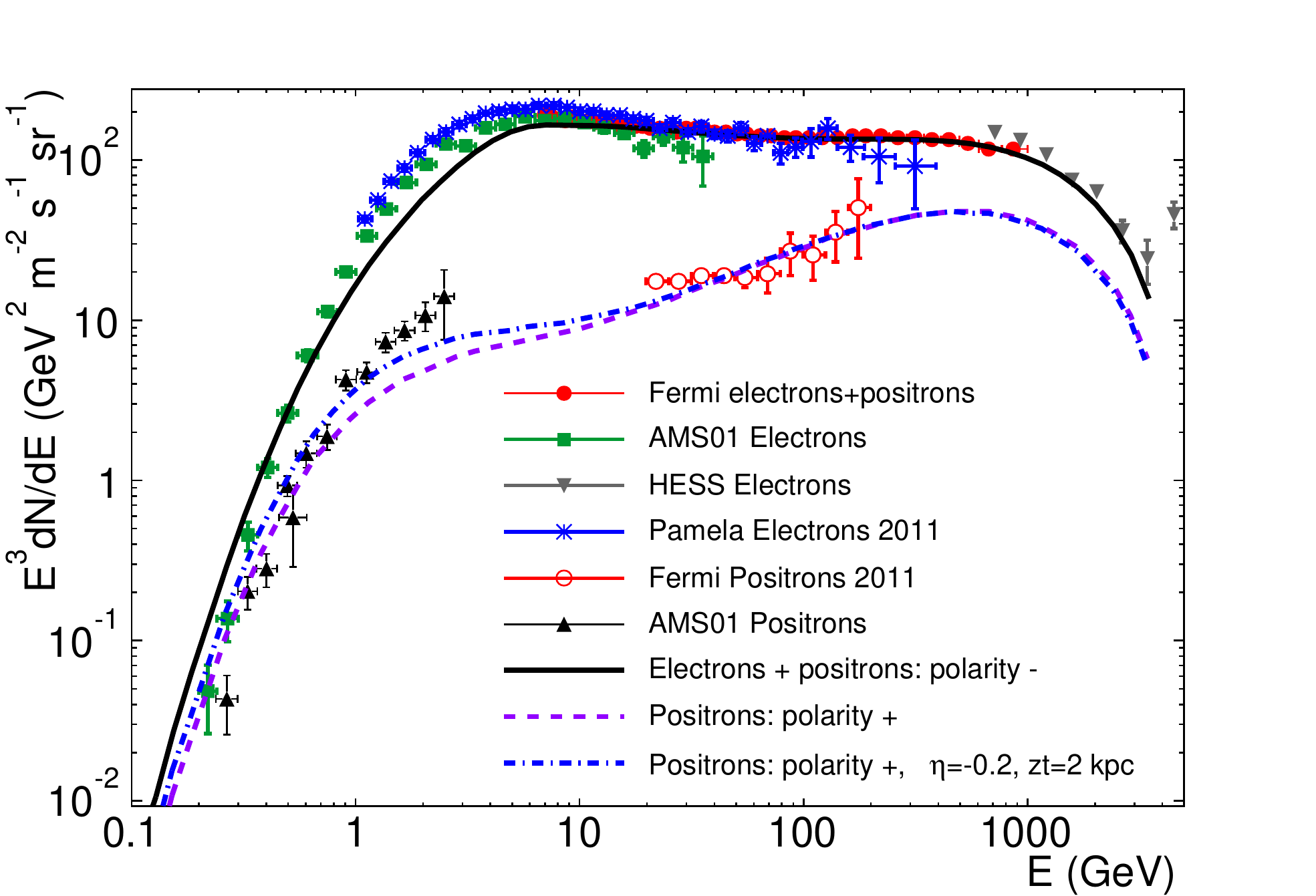}
\caption{The absolute $e^{-}$ and $e^{+}$ spectra measured by different experiments are compared with our calculations.}
\label{fig:PFfluxes}
\end{center}
\end{figure}

We show finally in Fig.~\ref{fig:PFfluxes} the absolute $e^{-}$ and $e^{+}$ spectra. 
%
%
We achieve a rather good agreement, within experimental uncertainties, with the observed absolute $e^{+}$ spectrum, which is however slightly underestimated between 1 and 3 GeV.  
%
The $e^{-}$ spectrum below 5 GeV is underestimated. In order to better understand this issue, we consider a modified model, in which we lower $z_{t}=2~\kpc$ and set $\eta=-0.2$. In order to ensure that the B/C ratio is still reproduced, we keep $D_{0}/z_{t}$ constant \cite{Berezinsky_book} and slightly readjust $v_{A}=13~\km/\s$. With these settings, we still correctly reproduce the B/C, $p$ and $\bar{p}$ data and affect only marginally the PF, but we improve the $e^{+}$ and the $e^{-}$ spectra. 
This may hint at low-energy modified diffusion effects being indeed relevant, but we must remark that theoretical uncertainties on secondary $e^{\pm}$ fluxes are large in this energy range because of poor knowledge of the production cross sections in $pp$ collisions \cite{Delahaye:2008ua,Delahaye:2010ji}.


{\em Conclusions}: Accounting for CR propagation in the solar system beyond the force-field description, is very important to understand present-day observations. We have computed the local interstellar spectra of several CR species and propagated them in the heliosphere by accounting for charge-sign dependent drifts. We applied our solar model to the B/C for the first time and showed that ACE observations can be interpreted within galactic propagation models in which also the observed spectra of $p$, He and $\bar{p}$ can be successfully reproduced without requiring spectral breaks. 

We then computed the CR $e^{-}$ and $e^{+}$ spectra and the positron fraction within the same models. The discordant observations at low energy by PAMELA and AMS-01 can be explained by charge-sign dependent drifts. For the first time, this result is achieved within galactic propagation models which are self-consistently derived from observations of CR nuclei, and by solely adjusting the solar propagation model to match the actual observation conditions of each experiment. 

The $e^{-}$ and $e^{+}$ low energy spectra are not satisfactorily reproduced, unless modified low-energy diffusion and a rather small halo are assumed. Enhancing the production of secondary leptons might improve the $e^{+}$ spectra, but has an impact on the $e^{-}$ spectrum only at very low energies. This might hint at the need of a more elaborated description of galactic and solar propagation, or of a more detailed analysis of secondary production cross sections. High quality and high statistics AMS-02 \cite{ams02} data, which are taken in a period of high solar activity and changing polarity, will help clarifying these issues.

We have shown results only for one particular set of propagation parameters. In fact, we tested several combinations of solar and galactic parameters, finding that in many cases a good agreement with all datasets can be achieved. A more detailed study of the parameter space is left for future work.

{\em Acknowledgments:} We warmly thank P.~Ullio for encouraging discussions. We also thank G.~Di Bernardo, F.~Donato, C.~Evoli, N.~Fornengo, D.~Gaggero, D.~Grasso and G.~Raffelt for many insightful discussions and comments and for reading a preliminary version of this Letter. We thank A.~Kopp for help with the early stage of development of \SolarProp. Support from the AvH Foundation is acknowledged. 

\bibliography{paper_modello}

\begin{thebibliography}{45}%
\makeatletter
\providecommand \@ifxundefined [1]{%
 \@ifx{#1\undefined}
}%
\providecommand \@ifnum [1]{%
 \ifnum #1\expandafter \@firstoftwo
 \else \expandafter \@secondoftwo
 \fi
}%
\providecommand \@ifx [1]{%
 \ifx #1\expandafter \@firstoftwo
 \else \expandafter \@secondoftwo
 \fi
}%
\providecommand \natexlab [1]{#1}%
\providecommand \enquote  [1]{``#1''}%
\providecommand \bibnamefont  [1]{#1}%
\providecommand \bibfnamefont [1]{#1}%
\providecommand \citenamefont [1]{#1}%
\providecommand \href@noop [0]{\@secondoftwo}%
\providecommand \href [0]{\begingroup \@sanitize@url \@href}%
\providecommand \@href[1]{\@@startlink{#1}\@@href}%
\providecommand \@@href[1]{\endgroup#1\@@endlink}%
\providecommand \@sanitize@url [0]{\catcode `\\12\catcode `\$12\catcode
  `\&12\catcode `\#12\catcode `\^12\catcode `\_12\catcode `\%12\relax}%
\providecommand \@@startlink[1]{}%
\providecommand \@@endlink[0]{}%
\providecommand \url  [0]{\begingroup\@sanitize@url \@url }%
\providecommand \@url [1]{\endgroup\@href {#1}{\urlprefix }}%
\providecommand \urlprefix  [0]{URL }%
\providecommand \Eprint [0]{\href }%
\providecommand \doibase [0]{http://dx.doi.org/}%
\providecommand \selectlanguage [0]{\@gobble}%
\providecommand \bibinfo  [0]{\@secondoftwo}%
\providecommand \bibfield  [0]{\@secondoftwo}%
\providecommand \translation [1]{[#1]}%
\providecommand \BibitemOpen [0]{}%
\providecommand \bibitemStop [0]{}%
\providecommand \bibitemNoStop [0]{.\EOS\space}%
\providecommand \EOS [0]{\spacefactor3000\relax}%
\providecommand \BibitemShut  [1]{\csname bibitem#1\endcsname}%
\let\auto@bib@innerbib\@empty
\bibitem [{\citenamefont {Adriani}\ \emph {et~al.}(2009)\citenamefont {Adriani}
  \emph {et~al.}}]{Adriani:2008zr}%
  \BibitemOpen
  \bibfield  {author} {\bibinfo {author} {\bibfnamefont {O.}~\bibnamefont
  {Adriani}} \emph {et~al.} (\bibinfo {collaboration} {PAMELA Collaboration}),\
  }\href {\doibase 10.1038/nature07942} {\bibfield  {journal} {\bibinfo
  {journal} {Nature}\ }\textbf {\bibinfo {volume} {458}},\ \bibinfo {pages}
  {607} (\bibinfo {year} {2009})},\ \Eprint {http://arxiv.org/abs/0810.4995}
  {arXiv:0810.4995 [astro-ph]} \BibitemShut {NoStop}%
\bibitem [{\citenamefont {Ackermann}\ \emph {et~al.}(2012)\citenamefont
  {Ackermann} \emph {et~al.}}]{FermiLAT:2011ab}%
  \BibitemOpen
  \bibfield  {author} {\bibinfo {author} {\bibfnamefont {M.}~\bibnamefont
  {Ackermann}} \emph {et~al.} (\bibinfo {collaboration} {Fermi LAT
  Collaboration}),\ }\href {\doibase 10.1103/PhysRevLett.108.011103} {\bibfield
   {journal} {\bibinfo  {journal} {Phys.Rev.Lett.}\ }\textbf {\bibinfo {volume}
  {108}},\ \bibinfo {pages} {011103} (\bibinfo {year} {2012})},\ \Eprint
  {http://arxiv.org/abs/1109.0521} {arXiv:1109.0521 [astro-ph.HE]} \BibitemShut
  {NoStop}%
\bibitem [{\citenamefont {Abdo}\ \emph {et~al.}(2009)\citenamefont {Abdo} \emph
  {et~al.}}]{Abdo:2009zk}%
  \BibitemOpen
  \bibfield  {author} {\bibinfo {author} {\bibfnamefont {A.~A.}\ \bibnamefont
  {Abdo}} \emph {et~al.} (\bibinfo {collaboration} {Fermi LAT Collaboration}),\
  }\href {\doibase 10.1103/PhysRevLett.102.181101} {\bibfield  {journal}
  {\bibinfo  {journal} {Phys.Rev.Lett.}\ }\textbf {\bibinfo {volume} {102}},\
  \bibinfo {pages} {181101} (\bibinfo {year} {2009})},\ \Eprint
  {http://arxiv.org/abs/0905.0025} {arXiv:0905.0025 [astro-ph.HE]} \BibitemShut
  {NoStop}%
\bibitem [{\citenamefont {Ackermann}\ \emph {et~al.}(2010)\citenamefont
  {Ackermann} \emph {et~al.}}]{Ackermann:2010ij}%
  \BibitemOpen
  \bibfield  {author} {\bibinfo {author} {\bibfnamefont {M.}~\bibnamefont
  {Ackermann}} \emph {et~al.} (\bibinfo {collaboration} {Fermi LAT
  Collaboration}),\ }\href {\doibase 10.1103/PhysRevD.82.092004} {\bibfield
  {journal} {\bibinfo  {journal} {Phys.Rev.}\ }\textbf {\bibinfo {volume}
  {D82}},\ \bibinfo {pages} {092004} (\bibinfo {year} {2010})},\ \Eprint
  {http://arxiv.org/abs/1008.3999} {arXiv:1008.3999 [astro-ph.HE]} \BibitemShut
  {NoStop}%
\bibitem [{\citenamefont {Aguilar}\ \emph {et~al.}(2007)\citenamefont {Aguilar}
  \emph {et~al.}}]{Aguilar:2007yf}%
  \BibitemOpen
  \bibfield  {author} {\bibinfo {author} {\bibfnamefont {M.}~\bibnamefont
  {Aguilar}} \emph {et~al.} (\bibinfo {collaboration} {AMS-01 Collaboration}),\
  }\href {\doibase 10.1016/j.physletb.2007.01.024} {\bibfield  {journal}
  {\bibinfo  {journal} {Phys.Lett.}\ }\textbf {\bibinfo {volume} {B646}},\
  \bibinfo {pages} {145} (\bibinfo {year} {2007})},\ \Eprint
  {http://arxiv.org/abs/astro-ph/0703154} {arXiv:astro-ph/0703154 [ASTRO-PH]}
  \BibitemShut {NoStop}%
\bibitem [{\citenamefont {Delahaye}\ \emph {et~al.}(2009)\citenamefont
  {Delahaye}, \citenamefont {Donato}, \citenamefont {Fornengo}, \citenamefont
  {Lavalle}, \citenamefont {Lineros} \emph {et~al.}}]{Delahaye:2008ua}%
  \BibitemOpen
  \bibfield  {author} {\bibinfo {author} {\bibfnamefont {T.}~\bibnamefont
  {Delahaye}}, \bibinfo {author} {\bibfnamefont {F.}~\bibnamefont {Donato}},
  \bibinfo {author} {\bibfnamefont {N.}~\bibnamefont {Fornengo}}, \bibinfo
  {author} {\bibfnamefont {J.}~\bibnamefont {Lavalle}}, \bibinfo {author}
  {\bibfnamefont {R.}~\bibnamefont {Lineros}},  \emph {et~al.},\ }\href
  {\doibase 10.1051/0004-6361/200811130} {\bibfield  {journal} {\bibinfo
  {journal} {Astron.Astrophys.}\ }\textbf {\bibinfo {volume} {501}},\ \bibinfo
  {pages} {821} (\bibinfo {year} {2009})},\ \Eprint
  {http://arxiv.org/abs/0809.5268} {arXiv:0809.5268 [astro-ph]} \BibitemShut
  {NoStop}%
\bibitem [{\citenamefont {Gast}\ and\ \citenamefont
  {Schael}(2009)}]{GastSchael}%
  \BibitemOpen
  \bibfield  {author} {\bibinfo {author} {\bibfnamefont {H.}~\bibnamefont
  {Gast}}\ and\ \bibinfo {author} {\bibfnamefont {S.}~\bibnamefont {Schael}}\
  }(\bibinfo {year} {2009})\ \bibinfo {note} {proceedings of the 31st ICRC,
  Lodz}\BibitemShut {NoStop}%
\bibitem [{\citenamefont {Della~Torre}\ \emph {et~al.}(2012)\citenamefont
  {Della~Torre}, \citenamefont {Bobik}, \citenamefont {Boschini}, \citenamefont
  {Consolandi}, \citenamefont {Gervasi} \emph {et~al.}}]{DellaTorre:2012zz}%
  \BibitemOpen
  \bibfield  {author} {\bibinfo {author} {\bibfnamefont {S.}~\bibnamefont
  {Della~Torre}}, \bibinfo {author} {\bibfnamefont {P.}~\bibnamefont {Bobik}},
  \bibinfo {author} {\bibfnamefont {M.~J.}\ \bibnamefont {Boschini}}, \bibinfo
  {author} {\bibfnamefont {C.}~\bibnamefont {Consolandi}}, \bibinfo {author}
  {\bibfnamefont {M.}~\bibnamefont {Gervasi}},  \emph {et~al.},\ }\href
  {\doibase 10.1016/j.asr.2012.02.017} {\bibfield  {journal} {\bibinfo
  {journal} {Adv.Space Res.}\ }\textbf {\bibinfo {volume} {49}},\ \bibinfo
  {pages} {1587} (\bibinfo {year} {2012})}\BibitemShut {NoStop}%
\bibitem [{\citenamefont {Delahaye}\ \emph {et~al.}(2010)\citenamefont
  {Delahaye}, \citenamefont {Lavalle}, \citenamefont {Lineros}, \citenamefont
  {Donato},\ and\ \citenamefont {Fornengo}}]{Delahaye:2010ji}%
  \BibitemOpen
  \bibfield  {author} {\bibinfo {author} {\bibfnamefont {T.}~\bibnamefont
  {Delahaye}}, \bibinfo {author} {\bibfnamefont {J.}~\bibnamefont {Lavalle}},
  \bibinfo {author} {\bibfnamefont {R.}~\bibnamefont {Lineros}}, \bibinfo
  {author} {\bibfnamefont {F.}~\bibnamefont {Donato}}, \ and\ \bibinfo {author}
  {\bibfnamefont {N.}~\bibnamefont {Fornengo}},\ }\href {\doibase
  10.1051/0004-6361/201014225} {\bibfield  {journal} {\bibinfo  {journal}
  {Astron.Astrophys.}\ }\textbf {\bibinfo {volume} {524}},\ \bibinfo {pages}
  {A51} (\bibinfo {year} {2010})},\ \Eprint {http://arxiv.org/abs/1002.1910}
  {arXiv:1002.1910 [astro-ph.HE]} \BibitemShut {NoStop}%
\bibitem [{\citenamefont {{Gleeson}}\ and\ \citenamefont
  {{Axford}}(1968)}]{Gleeson_1968ApJ}%
  \BibitemOpen
  \bibfield  {author} {\bibinfo {author} {\bibfnamefont {L.~J.}\ \bibnamefont
  {{Gleeson}}}\ and\ \bibinfo {author} {\bibfnamefont {W.~I.}\ \bibnamefont
  {{Axford}}},\ }\href {\doibase 10.1086/149822} {\bibfield  {journal}
  {\bibinfo  {journal} {\apj}\ }\textbf {\bibinfo {volume} {154}},\ \bibinfo
  {pages} {1011} (\bibinfo {year} {1968})}\BibitemShut {NoStop}%
\bibitem [{\citenamefont {{Clem}}\ \emph {et~al.}(1996)\citenamefont {{Clem}},
  \citenamefont {{Clements}}, \citenamefont {{Esposito}}, \citenamefont
  {{Evenson}}, \citenamefont {{Huber}}, \citenamefont {{L'Heureux}},
  \citenamefont {{Meyer}},\ and\ \citenamefont
  {{Constantin}}}]{1996ApJ...464..507C}%
  \BibitemOpen
  \bibfield  {author} {\bibinfo {author} {\bibfnamefont {J.~M.}\ \bibnamefont
  {{Clem}}}, \bibinfo {author} {\bibfnamefont {D.~P.}\ \bibnamefont
  {{Clements}}}, \bibinfo {author} {\bibfnamefont {J.}~\bibnamefont
  {{Esposito}}}, \bibinfo {author} {\bibfnamefont {P.}~\bibnamefont
  {{Evenson}}}, \bibinfo {author} {\bibfnamefont {D.}~\bibnamefont {{Huber}}},
  \bibinfo {author} {\bibfnamefont {J.}~\bibnamefont {{L'Heureux}}}, \bibinfo
  {author} {\bibfnamefont {P.}~\bibnamefont {{Meyer}}}, \ and\ \bibinfo
  {author} {\bibfnamefont {C.}~\bibnamefont {{Constantin}}},\ }\href {\doibase
  10.1086/177340} {\bibfield  {journal} {\bibinfo  {journal} {\apj}\ }\textbf
  {\bibinfo {volume} {464}},\ \bibinfo {pages} {507} (\bibinfo {year}
  {1996})}\BibitemShut {NoStop}%
\bibitem [{wil()}]{wilcox}%
  \BibitemOpen
  \href@noop {} {}\bibinfo {howpublished}
  {\url{http://wso.stanford.edu/}}\BibitemShut {NoStop}%
\bibitem [{\citenamefont {{Burlaga}}\ \emph {et~al.}(1981)\citenamefont
  {{Burlaga}}, \citenamefont {{Hundhausen}},\ and\ \citenamefont
  {{Zhao}}}]{1981JGR....86.8893B}%
  \BibitemOpen
  \bibfield  {author} {\bibinfo {author} {\bibfnamefont {L.~F.}\ \bibnamefont
  {{Burlaga}}}, \bibinfo {author} {\bibfnamefont {A.~J.}\ \bibnamefont
  {{Hundhausen}}}, \ and\ \bibinfo {author} {\bibfnamefont {X.-P.}\
  \bibnamefont {{Zhao}}},\ }\href {\doibase 10.1029/JA086iA11p08893} {\bibfield
   {journal} {\bibinfo  {journal} {\jgr}\ }\textbf {\bibinfo {volume} {86}},\
  \bibinfo {pages} {8893} (\bibinfo {year} {1981})}\BibitemShut {NoStop}%
\bibitem [{\citenamefont {Strauss}\ \emph {et~al.}(2012)\citenamefont
  {Strauss}, \citenamefont {Potgieter},\ and\ \citenamefont
  {Ferreira}}]{Strauss:2012zza}%
  \BibitemOpen
  \bibfield  {author} {\bibinfo {author} {\bibfnamefont {R.}~\bibnamefont
  {Strauss}}, \bibinfo {author} {\bibfnamefont {M.}~\bibnamefont {Potgieter}},
  \ and\ \bibinfo {author} {\bibfnamefont {S.}~\bibnamefont {Ferreira}},\
  }\href {\doibase 10.1016/j.asr.2011.10.006} {\bibfield  {journal} {\bibinfo
  {journal} {Adv.Space Res.}\ }\textbf {\bibinfo {volume} {49}},\ \bibinfo
  {pages} {392} (\bibinfo {year} {2012})}\BibitemShut {NoStop}%
\bibitem [{ams()}]{ams02}%
  \BibitemOpen
  \href@noop {} {}\bibinfo {howpublished}
  {\url{http://www.ams02.org}}\BibitemShut {NoStop}%
\bibitem [{\citenamefont {{Ginzburg}}\ and\ \citenamefont
  {{Syrovatskii}}(1964)}]{1964ocr..book.....G}%
  \BibitemOpen
  \bibfield  {author} {\bibinfo {author} {\bibfnamefont {V.~L.}\ \bibnamefont
  {{Ginzburg}}}\ and\ \bibinfo {author} {\bibfnamefont {S.~I.}\ \bibnamefont
  {{Syrovatskii}}},\ }\href@noop {} {\emph {\bibinfo {title} {The Origin of
  Cosmic Rays, New York: Macmillan, 1964}}}\ (\bibinfo {year}
  {1964})\BibitemShut {NoStop}%
\bibitem [{\citenamefont {{Parker}}(1965)}]{1965P&SS...13....9P}%
  \BibitemOpen
  \bibfield  {author} {\bibinfo {author} {\bibfnamefont {E.~N.}\ \bibnamefont
  {{Parker}}},\ }\href {\doibase 10.1016/0032-0633(65)90131-5} {\bibfield
  {journal} {\bibinfo  {journal} {\planss}\ }\textbf {\bibinfo {volume} {13}},\
  \bibinfo {pages} {9} (\bibinfo {year} {1965})}\BibitemShut {NoStop}%
\bibitem [{dra()}]{dragonweb}%
  \BibitemOpen
  \href@noop {} {}\bibinfo {howpublished}
  {\url{http://dragon.hepforge.org/}}\BibitemShut {NoStop}%
\bibitem [{\citenamefont {Bringmann}\ \emph {et~al.}(2012)\citenamefont
  {Bringmann}, \citenamefont {Donato},\ and\ \citenamefont
  {Lineros}}]{Bringmann:2011py}%
  \BibitemOpen
  \bibfield  {author} {\bibinfo {author} {\bibfnamefont {T.}~\bibnamefont
  {Bringmann}}, \bibinfo {author} {\bibfnamefont {F.}~\bibnamefont {Donato}}, \
  and\ \bibinfo {author} {\bibfnamefont {R.~A.}\ \bibnamefont {Lineros}},\
  }\href {\doibase 10.1088/1475-7516/2012/01/049} {\bibfield  {journal}
  {\bibinfo  {journal} {JCAP}\ }\textbf {\bibinfo {volume} {1201}},\ \bibinfo
  {pages} {049} (\bibinfo {year} {2012})},\ \Eprint
  {http://arxiv.org/abs/1106.4821} {arXiv:1106.4821 [astro-ph.GA]} \BibitemShut
  {NoStop}%
\bibitem [{\citenamefont {Strong}\ \emph {et~al.}(2011)\citenamefont {Strong},
  \citenamefont {Orlando},\ and\ \citenamefont {Jaffe}}]{Strong:2011wd}%
  \BibitemOpen
  \bibfield  {author} {\bibinfo {author} {\bibfnamefont {A.}~\bibnamefont
  {Strong}}, \bibinfo {author} {\bibfnamefont {E.}~\bibnamefont {Orlando}}, \
  and\ \bibinfo {author} {\bibfnamefont {T.}~\bibnamefont {Jaffe}},\
  }\href@noop {} {\bibfield  {journal} {\bibinfo  {journal}
  {Astron.Astrophys.}\ }\textbf {\bibinfo {volume} {534}},\ \bibinfo {pages}
  {A54} (\bibinfo {year} {2011})},\ \Eprint {http://arxiv.org/abs/1108.4822}
  {arXiv:1108.4822 [astro-ph.HE]} \BibitemShut {NoStop}%
\bibitem [{\citenamefont {Di~Bernardo}\ \emph {et~al.}(2012)\citenamefont
  {Di~Bernardo}, \citenamefont {Evoli}, \citenamefont {Gaggero}, \citenamefont
  {Grasso},\ and\ \citenamefont {Maccione}}]{DiBernardo:2012zu}%
  \BibitemOpen
  \bibfield  {author} {\bibinfo {author} {\bibfnamefont {G.}~\bibnamefont
  {Di~Bernardo}}, \bibinfo {author} {\bibfnamefont {C.}~\bibnamefont {Evoli}},
  \bibinfo {author} {\bibfnamefont {D.}~\bibnamefont {Gaggero}}, \bibinfo
  {author} {\bibfnamefont {D.}~\bibnamefont {Grasso}}, \ and\ \bibinfo {author}
  {\bibfnamefont {L.}~\bibnamefont {Maccione}},\ }\href@noop {} {\  (\bibinfo
  {year} {2012})},\ \Eprint {http://arxiv.org/abs/1210.4546} {arXiv:1210.4546
  [astro-ph.HE]} \BibitemShut {NoStop}%
\bibitem [{\citenamefont {Grasso}\ \emph {et~al.}(2009)\citenamefont {Grasso}
  \emph {et~al.}}]{Grasso:2009ma}%
  \BibitemOpen
  \bibfield  {author} {\bibinfo {author} {\bibfnamefont {D.}~\bibnamefont
  {Grasso}} \emph {et~al.} (\bibinfo {collaboration} {FERMI-LAT
  Collaboration}),\ }\href {\doibase 10.1016/j.astropartphys.2009.07.003}
  {\bibfield  {journal} {\bibinfo  {journal} {Astropart.Phys.}\ }\textbf
  {\bibinfo {volume} {32}},\ \bibinfo {pages} {140} (\bibinfo {year} {2009})},\
  \Eprint {http://arxiv.org/abs/0905.0636} {arXiv:0905.0636 [astro-ph.HE]}
  \BibitemShut {NoStop}%
\bibitem [{\citenamefont {Di~Bernardo}\ \emph {et~al.}(2011)\citenamefont
  {Di~Bernardo}, \citenamefont {Evoli}, \citenamefont {Gaggero}, \citenamefont
  {Grasso}, \citenamefont {Maccione} \emph {et~al.}}]{DiBernardo:2010is}%
  \BibitemOpen
  \bibfield  {author} {\bibinfo {author} {\bibfnamefont {G.}~\bibnamefont
  {Di~Bernardo}}, \bibinfo {author} {\bibfnamefont {C.}~\bibnamefont {Evoli}},
  \bibinfo {author} {\bibfnamefont {D.}~\bibnamefont {Gaggero}}, \bibinfo
  {author} {\bibfnamefont {D.}~\bibnamefont {Grasso}}, \bibinfo {author}
  {\bibfnamefont {L.}~\bibnamefont {Maccione}},  \emph {et~al.},\ }\href
  {\doibase 10.1016/j.astropartphys.2010.11.005} {\bibfield  {journal}
  {\bibinfo  {journal} {Astropart.Phys.}\ }\textbf {\bibinfo {volume} {34}},\
  \bibinfo {pages} {528} (\bibinfo {year} {2011})},\ \Eprint
  {http://arxiv.org/abs/1010.0174} {arXiv:1010.0174 [astro-ph.HE]} \BibitemShut
  {NoStop}%
\bibitem [{\citenamefont {{Strauss}}\ \emph {et~al.}(2011)\citenamefont
  {{Strauss}}, \citenamefont {{Potgieter}}, \citenamefont {{B{\"u}sching}},\
  and\ \citenamefont {{Kopp}}}]{2011ApJ...735...83S}%
  \BibitemOpen
  \bibfield  {author} {\bibinfo {author} {\bibfnamefont {R.~D.}\ \bibnamefont
  {{Strauss}}}, \bibinfo {author} {\bibfnamefont {M.~S.}\ \bibnamefont
  {{Potgieter}}}, \bibinfo {author} {\bibfnamefont {I.}~\bibnamefont
  {{B{\"u}sching}}}, \ and\ \bibinfo {author} {\bibfnamefont {A.}~\bibnamefont
  {{Kopp}}},\ }\href {\doibase 10.1088/0004-637X/735/2/83} {\bibfield
  {journal} {\bibinfo  {journal} {\apj}\ }\textbf {\bibinfo {volume} {735}},\
  \bibinfo {eid} {83} (\bibinfo {year} {2011})}\BibitemShut {NoStop}%
\bibitem [{\citenamefont {{Strauss}}\ \emph {et~al.}(2012)\citenamefont
  {{Strauss}}, \citenamefont {{Potgieter}}, \citenamefont {{B{\"u}sching}},\
  and\ \citenamefont {{Kopp}}}]{2012Ap&SS.339..223S}%
  \BibitemOpen
  \bibfield  {author} {\bibinfo {author} {\bibfnamefont {R.~D.}\ \bibnamefont
  {{Strauss}}}, \bibinfo {author} {\bibfnamefont {M.~S.}\ \bibnamefont
  {{Potgieter}}}, \bibinfo {author} {\bibfnamefont {I.}~\bibnamefont
  {{B{\"u}sching}}}, \ and\ \bibinfo {author} {\bibfnamefont {A.}~\bibnamefont
  {{Kopp}}},\ }\href {\doibase 10.1007/s10509-012-1003-z} {\bibfield  {journal}
  {\bibinfo  {journal} {\apss}\ }\textbf {\bibinfo {volume} {339}},\ \bibinfo
  {pages} {223} (\bibinfo {year} {2012})}\BibitemShut {NoStop}%
\bibitem [{\citenamefont {{Alanko-Huotari}}\ \emph {et~al.}(2007)\citenamefont
  {{Alanko-Huotari}}, \citenamefont {{Usoskin}}, \citenamefont {{Mursula}},\
  and\ \citenamefont {{Kovaltsov}}}]{2007JGRA..11208101A}%
  \BibitemOpen
  \bibfield  {author} {\bibinfo {author} {\bibfnamefont {K.}~\bibnamefont
  {{Alanko-Huotari}}}, \bibinfo {author} {\bibfnamefont {I.~G.}\ \bibnamefont
  {{Usoskin}}}, \bibinfo {author} {\bibfnamefont {K.}~\bibnamefont
  {{Mursula}}}, \ and\ \bibinfo {author} {\bibfnamefont {G.~A.}\ \bibnamefont
  {{Kovaltsov}}},\ }\href {\doibase 10.1029/2007JA012280} {\bibfield  {journal}
  {\bibinfo  {journal} {Journal of Geophysical Research (Space Physics)}\
  }\textbf {\bibinfo {volume} {112}},\ \bibinfo {eid} {A08101} (\bibinfo {year}
  {2007})}\BibitemShut {NoStop}%
\bibitem [{\citenamefont {Gardiner}(2009)}]{gardiner2009stochastic}%
  \BibitemOpen
  \bibfield  {author} {\bibinfo {author} {\bibfnamefont {C.}~\bibnamefont
  {Gardiner}},\ }\href {http://books.google.de/books?id=otg3PQAACAAJ} {\emph
  {\bibinfo {title} {Stochastic Methods: A Handbook for the Natural and Social
  Sciences}}},\ Springer Series in Synergetics\ (\bibinfo  {publisher}
  {Springer},\ \bibinfo {year} {2009})\BibitemShut {NoStop}%
\bibitem [{\citenamefont {{Kopp}}\ \emph {et~al.}(2012)\citenamefont {{Kopp}},
  \citenamefont {{B{\"u}sching}}, \citenamefont {{Strauss}},\ and\
  \citenamefont {{Potgieter}}}]{2012CoPhC.183..530K}%
  \BibitemOpen
  \bibfield  {author} {\bibinfo {author} {\bibfnamefont {A.}~\bibnamefont
  {{Kopp}}}, \bibinfo {author} {\bibfnamefont {I.}~\bibnamefont
  {{B{\"u}sching}}}, \bibinfo {author} {\bibfnamefont {R.~D.}\ \bibnamefont
  {{Strauss}}}, \ and\ \bibinfo {author} {\bibfnamefont {M.~S.}\ \bibnamefont
  {{Potgieter}}},\ }\href {\doibase 10.1016/j.cpc.2011.11.014} {\bibfield
  {journal} {\bibinfo  {journal} {Computer Physics Communications}\ }\textbf
  {\bibinfo {volume} {183}},\ \bibinfo {pages} {530} (\bibinfo {year}
  {2012})}\BibitemShut {NoStop}%
\bibitem [{\citenamefont {Bobik}\ \emph {et~al.}(2012)\citenamefont {Bobik},
  \citenamefont {Boella}, \citenamefont {Boschini}, \citenamefont {Consolandi},
  \citenamefont {Della~Torre} \emph {et~al.}}]{Bobik:2011ig}%
  \BibitemOpen
  \bibfield  {author} {\bibinfo {author} {\bibfnamefont {P.}~\bibnamefont
  {Bobik}}, \bibinfo {author} {\bibfnamefont {G.}~\bibnamefont {Boella}},
  \bibinfo {author} {\bibfnamefont {M.}~\bibnamefont {Boschini}}, \bibinfo
  {author} {\bibfnamefont {C.}~\bibnamefont {Consolandi}}, \bibinfo {author}
  {\bibfnamefont {S.}~\bibnamefont {Della~Torre}},  \emph {et~al.},\ }\href
  {\doibase 10.1088/0004-637X/745/2/132} {\bibfield  {journal} {\bibinfo
  {journal} {Astrophys.J.}\ }\textbf {\bibinfo {volume} {745}},\ \bibinfo
  {pages} {132} (\bibinfo {year} {2012})},\ \Eprint
  {http://arxiv.org/abs/1110.4315} {arXiv:1110.4315 [astro-ph.SR]} \BibitemShut
  {NoStop}%
\bibitem [{\citenamefont {Dr{\"{o}}ge}(2005)}]{Droge2005532}%
  \BibitemOpen
  \bibfield  {author} {\bibinfo {author} {\bibfnamefont {W.}~\bibnamefont
  {Dr{\"{o}}ge}},\ }\href {\doibase 10.1016/j.asr.2004.12.007} {\bibfield
  {journal} {\bibinfo  {journal} {Advances in Space Research}\ }\textbf
  {\bibinfo {volume} {35}},\ \bibinfo {pages} {532 } (\bibinfo {year}
  {2005})}\BibitemShut {NoStop}%
\bibitem [{\citenamefont {{Giacalone}}\ and\ \citenamefont
  {{Jokipii}}(1999)}]{1999ApJ...520..204G}%
  \BibitemOpen
  \bibfield  {author} {\bibinfo {author} {\bibfnamefont {J.}~\bibnamefont
  {{Giacalone}}}\ and\ \bibinfo {author} {\bibfnamefont {J.~R.}\ \bibnamefont
  {{Jokipii}}},\ }\href {\doibase 10.1086/307452} {\bibfield  {journal}
  {\bibinfo  {journal} {\apj}\ }\textbf {\bibinfo {volume} {520}},\ \bibinfo
  {pages} {204} (\bibinfo {year} {1999})}\BibitemShut {NoStop}%
\bibitem [{\citenamefont {{Jokipii}}\ and\ \citenamefont
  {{Levy}}(1977)}]{1977ApJ...213L..85J}%
  \BibitemOpen
  \bibfield  {author} {\bibinfo {author} {\bibfnamefont {J.~R.}\ \bibnamefont
  {{Jokipii}}}\ and\ \bibinfo {author} {\bibfnamefont {E.~H.}\ \bibnamefont
  {{Levy}}},\ }\href {\doibase 10.1086/182415} {\bibfield  {journal} {\bibinfo
  {journal} {\apjl}\ }\textbf {\bibinfo {volume} {213}},\ \bibinfo {pages}
  {L85} (\bibinfo {year} {1977})}\BibitemShut {NoStop}%
\bibitem [{\citenamefont {Ferreira}\ and\ \citenamefont
  {Potgieter}(2003)}]{Ferreira2003657}%
  \BibitemOpen
  \bibfield  {author} {\bibinfo {author} {\bibfnamefont {S.}~\bibnamefont
  {Ferreira}}\ and\ \bibinfo {author} {\bibfnamefont {M.}~\bibnamefont
  {Potgieter}},\ }\href {\doibase 10.1016/S0273-1177(03)00360-0} {\bibfield
  {journal} {\bibinfo  {journal} {Advances in Space Research}\ }\textbf
  {\bibinfo {volume} {32}},\ \bibinfo {pages} {657 } (\bibinfo {year}
  {2003})}\BibitemShut {NoStop}%
\bibitem [{\citenamefont {{Ferreira}}\ and\ \citenamefont
  {{Potgieter}}(2004)}]{2004ApJ...603..744F}%
  \BibitemOpen
  \bibfield  {author} {\bibinfo {author} {\bibfnamefont {S.~E.~S.}\
  \bibnamefont {{Ferreira}}}\ and\ \bibinfo {author} {\bibfnamefont {M.~S.}\
  \bibnamefont {{Potgieter}}},\ }\href {\doibase 10.1086/381649} {\bibfield
  {journal} {\bibinfo  {journal} {\apj}\ }\textbf {\bibinfo {volume} {603}},\
  \bibinfo {pages} {744} (\bibinfo {year} {2004})}\BibitemShut {NoStop}%
\bibitem [{\citenamefont {{George}}\ \emph {et~al.}(2009)\citenamefont
  {{George}}, \citenamefont {{Lave}}, \citenamefont {{Wiedenbeck}},
  \citenamefont {{Binns}}, \citenamefont {{Cummings}}, \citenamefont {{Davis}},
  \citenamefont {{de Nolfo}}, \citenamefont {{Hink}}, \citenamefont {{Israel}},
  \citenamefont {{Leske}}, \citenamefont {{Mewaldt}}, \citenamefont {{Scott}},
  \citenamefont {{Stone}}, \citenamefont {{von Rosenvinge}},\ and\
  \citenamefont {{Yanasak}}}]{2009ApJ...698.1666G}%
  \BibitemOpen
  \bibfield  {author} {\bibinfo {author} {\bibfnamefont {J.~S.}\ \bibnamefont
  {{George}}}, \bibinfo {author} {\bibfnamefont {K.~A.}\ \bibnamefont
  {{Lave}}}, \bibinfo {author} {\bibfnamefont {M.~E.}\ \bibnamefont
  {{Wiedenbeck}}}, \bibinfo {author} {\bibfnamefont {W.~R.}\ \bibnamefont
  {{Binns}}}, \bibinfo {author} {\bibfnamefont {A.~C.}\ \bibnamefont
  {{Cummings}}}, \bibinfo {author} {\bibfnamefont {A.~J.}\ \bibnamefont
  {{Davis}}}, \bibinfo {author} {\bibfnamefont {G.~A.}\ \bibnamefont {{de
  Nolfo}}}, \bibinfo {author} {\bibfnamefont {P.~L.}\ \bibnamefont {{Hink}}},
  \bibinfo {author} {\bibfnamefont {M.~H.}\ \bibnamefont {{Israel}}}, \bibinfo
  {author} {\bibfnamefont {R.~A.}\ \bibnamefont {{Leske}}}, \bibinfo {author}
  {\bibfnamefont {R.~A.}\ \bibnamefont {{Mewaldt}}}, \bibinfo {author}
  {\bibfnamefont {L.~M.}\ \bibnamefont {{Scott}}}, \bibinfo {author}
  {\bibfnamefont {E.~C.}\ \bibnamefont {{Stone}}}, \bibinfo {author}
  {\bibfnamefont {T.~T.}\ \bibnamefont {{von Rosenvinge}}}, \ and\ \bibinfo
  {author} {\bibfnamefont {N.~E.}\ \bibnamefont {{Yanasak}}},\ }\href {\doibase
  10.1088/0004-637X/698/2/1666} {\bibfield  {journal} {\bibinfo  {journal}
  {\apj}\ }\textbf {\bibinfo {volume} {698}},\ \bibinfo {pages} {1666}
  (\bibinfo {year} {2009})}\BibitemShut {NoStop}%
\bibitem [{\citenamefont {{Binns}}\ \emph {et~al.}(1989)\citenamefont
  {{Binns}}, \citenamefont {{Garrard}}, \citenamefont {{Gibner}}, \citenamefont
  {{Israel}}, \citenamefont {{Kertzman}}, \citenamefont {{Klarmann}},
  \citenamefont {{Newport}}, \citenamefont {{Stone}},\ and\ \citenamefont
  {{Waddington}}}]{1989ApJ...346..997B}%
  \BibitemOpen
  \bibfield  {author} {\bibinfo {author} {\bibfnamefont {W.~R.}\ \bibnamefont
  {{Binns}}}, \bibinfo {author} {\bibfnamefont {T.~L.}\ \bibnamefont
  {{Garrard}}}, \bibinfo {author} {\bibfnamefont {P.~S.}\ \bibnamefont
  {{Gibner}}}, \bibinfo {author} {\bibfnamefont {M.~H.}\ \bibnamefont
  {{Israel}}}, \bibinfo {author} {\bibfnamefont {M.~P.}\ \bibnamefont
  {{Kertzman}}}, \bibinfo {author} {\bibfnamefont {J.}~\bibnamefont
  {{Klarmann}}}, \bibinfo {author} {\bibfnamefont {B.~J.}\ \bibnamefont
  {{Newport}}}, \bibinfo {author} {\bibfnamefont {E.~C.}\ \bibnamefont
  {{Stone}}}, \ and\ \bibinfo {author} {\bibfnamefont {C.~J.}\ \bibnamefont
  {{Waddington}}},\ }\href {\doibase 10.1086/168082} {\bibfield  {journal}
  {\bibinfo  {journal} {\apj}\ }\textbf {\bibinfo {volume} {346}},\ \bibinfo
  {pages} {997} (\bibinfo {year} {1989})}\BibitemShut {NoStop}%
\bibitem [{\citenamefont {{Swordy}}\ \emph {et~al.}(1990)\citenamefont
  {{Swordy}}, \citenamefont {{Mueller}}, \citenamefont {{Meyer}}, \citenamefont
  {{L'Heureux}},\ and\ \citenamefont {{Grunsfeld}}}]{1990ApJ...349..625S}%
  \BibitemOpen
  \bibfield  {author} {\bibinfo {author} {\bibfnamefont {S.~P.}\ \bibnamefont
  {{Swordy}}}, \bibinfo {author} {\bibfnamefont {D.}~\bibnamefont {{Mueller}}},
  \bibinfo {author} {\bibfnamefont {P.}~\bibnamefont {{Meyer}}}, \bibinfo
  {author} {\bibfnamefont {J.}~\bibnamefont {{L'Heureux}}}, \ and\ \bibinfo
  {author} {\bibfnamefont {J.~M.}\ \bibnamefont {{Grunsfeld}}},\ }\href
  {\doibase 10.1086/168349} {\bibfield  {journal} {\bibinfo  {journal} {\apj}\
  }\textbf {\bibinfo {volume} {349}},\ \bibinfo {pages} {625} (\bibinfo {year}
  {1990})}\BibitemShut {NoStop}%
\bibitem [{\citenamefont {Panov}\ \emph {et~al.}(2007)\citenamefont {Panov},
  \citenamefont {Sokolskaya}, \citenamefont {Adams}, \citenamefont {Ahn},
  \citenamefont {Bashindzhagyan} \emph {et~al.}}]{Panov:2007fe}%
  \BibitemOpen
  \bibfield  {author} {\bibinfo {author} {\bibfnamefont {A.}~\bibnamefont
  {Panov}}, \bibinfo {author} {\bibfnamefont {N.}~\bibnamefont {Sokolskaya}},
  \bibinfo {author} {\bibfnamefont {J.}~\bibnamefont {Adams}, \bibfnamefont
  {J.H.}}, \bibinfo {author} {\bibfnamefont {H.}~\bibnamefont {Ahn}}, \bibinfo
  {author} {\bibfnamefont {G.}~\bibnamefont {Bashindzhagyan}},  \emph
  {et~al.},\ }\href@noop {} {\  (\bibinfo {year} {2007})},\ \Eprint
  {http://arxiv.org/abs/0707.4415} {arXiv:0707.4415 [astro-ph]} \BibitemShut
  {NoStop}%
\bibitem [{\citenamefont {Ahn}\ \emph {et~al.}(2008)\citenamefont {Ahn},
  \citenamefont {Allison}, \citenamefont {Bagliesi}, \citenamefont {Beatty},
  \citenamefont {Bigongiari} \emph {et~al.}}]{Ahn:2008my}%
  \BibitemOpen
  \bibfield  {author} {\bibinfo {author} {\bibfnamefont {H.}~\bibnamefont
  {Ahn}}, \bibinfo {author} {\bibfnamefont {P.}~\bibnamefont {Allison}},
  \bibinfo {author} {\bibfnamefont {M.}~\bibnamefont {Bagliesi}}, \bibinfo
  {author} {\bibfnamefont {J.}~\bibnamefont {Beatty}}, \bibinfo {author}
  {\bibfnamefont {G.}~\bibnamefont {Bigongiari}},  \emph {et~al.},\ }\href
  {\doibase 10.1016/j.astropartphys.2008.07.010} {\bibfield  {journal}
  {\bibinfo  {journal} {Astropart.Phys.}\ }\textbf {\bibinfo {volume} {30}},\
  \bibinfo {pages} {133} (\bibinfo {year} {2008})},\ \Eprint
  {http://arxiv.org/abs/0808.1718} {arXiv:0808.1718 [astro-ph]} \BibitemShut
  {NoStop}%
\bibitem [{bes()}]{bess}%
  \BibitemOpen
  \href@noop {} {}\bibinfo {howpublished}
  {\url{http://bess.kek.jp}}\BibitemShut {NoStop}%
\bibitem [{\citenamefont {Di~Bernardo}\ \emph {et~al.}(2010)\citenamefont
  {Di~Bernardo}, \citenamefont {Evoli}, \citenamefont {Gaggero}, \citenamefont
  {Grasso},\ and\ \citenamefont {Maccione}}]{DiBernardo:2009ku}%
  \BibitemOpen
  \bibfield  {author} {\bibinfo {author} {\bibfnamefont {G.}~\bibnamefont
  {Di~Bernardo}}, \bibinfo {author} {\bibfnamefont {C.}~\bibnamefont {Evoli}},
  \bibinfo {author} {\bibfnamefont {D.}~\bibnamefont {Gaggero}}, \bibinfo
  {author} {\bibfnamefont {D.}~\bibnamefont {Grasso}}, \ and\ \bibinfo {author}
  {\bibfnamefont {L.}~\bibnamefont {Maccione}},\ }\href {\doibase
  10.1016/j.astropartphys.2010.08.006} {\bibfield  {journal} {\bibinfo
  {journal} {Astropart.Phys.}\ }\textbf {\bibinfo {volume} {34}},\ \bibinfo
  {pages} {274} (\bibinfo {year} {2010})},\ \Eprint
  {http://arxiv.org/abs/0909.4548} {arXiv:0909.4548 [astro-ph.HE]} \BibitemShut
  {NoStop}%
\bibitem [{\citenamefont {Putze}\ \emph {et~al.}(2010)\citenamefont {Putze},
  \citenamefont {Derome},\ and\ \citenamefont {Maurin}}]{Putze:2010zn}%
  \BibitemOpen
  \bibfield  {author} {\bibinfo {author} {\bibfnamefont {A.}~\bibnamefont
  {Putze}}, \bibinfo {author} {\bibfnamefont {L.}~\bibnamefont {Derome}}, \
  and\ \bibinfo {author} {\bibfnamefont {D.}~\bibnamefont {Maurin}},\
  }\href@noop {} {\bibfield  {journal} {\bibinfo  {journal}
  {Astron.Astrophys.}\ }\textbf {\bibinfo {volume} {516}},\ \bibinfo {pages}
  {A66} (\bibinfo {year} {2010})},\ \Eprint {http://arxiv.org/abs/1001.0551}
  {arXiv:1001.0551 [astro-ph.HE]} \BibitemShut {NoStop}%
\bibitem [{\citenamefont {Maurin}\ \emph {et~al.}(2010)\citenamefont {Maurin},
  \citenamefont {Putze},\ and\ \citenamefont {Derome}}]{Maurin:2010zp}%
  \BibitemOpen
  \bibfield  {author} {\bibinfo {author} {\bibfnamefont {D.}~\bibnamefont
  {Maurin}}, \bibinfo {author} {\bibfnamefont {A.}~\bibnamefont {Putze}}, \
  and\ \bibinfo {author} {\bibfnamefont {L.}~\bibnamefont {Derome}},\
  }\href@noop {} {\bibfield  {journal} {\bibinfo  {journal}
  {Astron.Astrophys.}\ }\textbf {\bibinfo {volume} {516}},\ \bibinfo {pages}
  {A67} (\bibinfo {year} {2010})},\ \Eprint {http://arxiv.org/abs/1001.0553}
  {arXiv:1001.0553 [astro-ph.HE]} \BibitemShut {NoStop}%
\bibitem [{\citenamefont {Trotta}\ \emph {et~al.}(2011)\citenamefont {Trotta},
  \citenamefont {Johannesson}, \citenamefont {Moskalenko}, \citenamefont
  {Porter}, \citenamefont {de~Austri} \emph {et~al.}}]{Trotta:2010mx}%
  \BibitemOpen
  \bibfield  {author} {\bibinfo {author} {\bibfnamefont {R.}~\bibnamefont
  {Trotta}}, \bibinfo {author} {\bibfnamefont {G.}~\bibnamefont {Johannesson}},
  \bibinfo {author} {\bibfnamefont {I.}~\bibnamefont {Moskalenko}}, \bibinfo
  {author} {\bibfnamefont {T.}~\bibnamefont {Porter}}, \bibinfo {author}
  {\bibfnamefont {R.~R.}\ \bibnamefont {de~Austri}},  \emph {et~al.},\ }\href
  {\doibase 10.1088/0004-637X/729/2/106} {\bibfield  {journal} {\bibinfo
  {journal} {Astrophys.J.}\ }\textbf {\bibinfo {volume} {729}},\ \bibinfo
  {pages} {106} (\bibinfo {year} {2011})},\ \Eprint
  {http://arxiv.org/abs/1011.0037} {arXiv:1011.0037 [astro-ph.HE]} \BibitemShut
  {NoStop}%
\bibitem [{\citenamefont {{Berezinskii}}\ \emph {et~al.}(1990)\citenamefont
  {{Berezinskii}}, \citenamefont {{Bulanov}}, \citenamefont {{Dogiel}},\ and\
  \citenamefont {{Ptuskin}}}]{Berezinsky_book}%
  \BibitemOpen
  \bibfield  {author} {\bibinfo {author} {\bibfnamefont {V.~S.}\ \bibnamefont
  {{Berezinskii}}}, \bibinfo {author} {\bibfnamefont {S.~V.}\ \bibnamefont
  {{Bulanov}}}, \bibinfo {author} {\bibfnamefont {V.~A.}\ \bibnamefont
  {{Dogiel}}}, \ and\ \bibinfo {author} {\bibfnamefont {V.~S.}\ \bibnamefont
  {{Ptuskin}}},\ }\href@noop {} {\emph {\bibinfo {title} {Amsterdam:
  North-Holland, 1990, edited by Ginzburg, V.L.}}},\ edited by\ \bibinfo
  {editor} {\bibnamefont {{Berezinskii, V.~S., Bulanov, S.~V., Dogiel, V.~A.,
  \& Ptuskin, V.~S. }}}\ (\bibinfo {year} {1990})\BibitemShut {NoStop}%
\end{thebibliography}%
\bibliographystyle{apsrev4-1}

\end{document}